\documentclass[doublecol]{epl2}
\usepackage{url}

\title{Imitate or innovate: competition of strategy updating attitudes in spatial social dilemma games}
\shorttitle{Imitate or innovate: competition of strategy updating attitudes}

\author{Zsuzsa Danku\inst{1}, Zhen Wang\inst{2}, and Attila Szolnoki\inst{3}}
\shortauthor{Danku, Wang, and Szolnoki}
\institute{\inst{1}Institute of Mathematics and Informatics, University of Ny{\'\i}regyh\'aza, Ny{\'\i}regyh\'aza, Hungary\\
\inst{2}School of Cyberspace, Hangzhou Dianzi University, Hangzhou
310018, China\\
\inst{3}Institute of Technical Physics and Materials Science, Centre for Energy Research, Hungarian Academy of Sciences, P.O. Box 49, H-1525 Budapest, Hungary\\}

\abstract{Evolution is based on the assumption that competing players update their strategies to increase their individual payoffs. However, while the applied updating method can be different, most of previous works proposed uniform models where players use identical way to revise their strategies. In this work we explore how imitation-based or learning attitude and innovation-based or myopic best response attitude compete for space in a complex model where both attitudes are available. In the absence of additional cost the best response trait practically dominates the whole snow-drift game parameter space which is in agreement with the average payoff difference of basic models. When additional cost is involved then the imitation attitude can gradually invade the whole parameter space but this transition happens in a highly nontrivial way. However, the role of competing attitudes is reversed in the stag-hunt parameter space where imitation is more successful in general. Interestingly, a four-state solution can be observed for the latter game which is a consequence of an emerging cyclic dominance between possible states. These phenomena can be understood by analyzing the microscopic invasion processes, which reveals the unequal propagation velocities of strategies and attitudes.}

\begin{document}

\maketitle
To imitate a more successful strategy is a frequently applied microscopic rule within the framework of evolutionary game theoretical models which focus on the fundamental conflict of individual and community benefits \cite{nowak_11,sigmund_10}. This assumption is partly motivated by biological systems where payoff is interpreted as fitness or reproductive success \cite{maynard_82}. Considering more sophisticated human systems, where similar social dilemmas are on stage, there are other alternative suggestions for strategy updating rules that take account of cognitive skills of competitors. During the last decades theoretical models have raised several ways how to update strategies including myopic best response \cite{matsui_jet92, blume_l_geb93, roca_epjb09, szabo_pre10, szabo_jtb12}, learning, or reinforcement learning strategies \cite{macy_pnas02,izquierdo_geb07,masuda_jtb11,cimini_jrsif14,li_k_epl16,horita_srep17,stivala_pre16,broere_srep17,takesue_epl17}. In parallel, a huge number of experimental works have been published, but sometimes their conclusions are conflicting which make difficult the comparison with theoretical predictions
\cite{traulsen_pnas10,gracia-lazaro_pnas12,rand_pnas14,tinghog_n13}.

One of the possible reasons of contradicting experimental results could be that we cannot be fully sure what is the microscopic motivation of individual competitors when they update their strategies. Furthermore the simultaneous presence of different updating traits or attitudes cannot be excluded, which makes the evaluation of different external conditions even harder. Interestingly, this fact has been largely ignored by theoretical works because most of them assume uniform players in the sense that they all apply the same method or attitude to revise their present states. In this letter we consider a simple model where two conceptually different attitudes are available for individuals who try to reach a higher payoff. These strategy updating methods are based on imitation or innovation and players are using one of them during a microscopic step. Beside heterogeneous attitudes we also extend the basic models by considering the fact that applying a certain attitude may be costly. For example, innovation requires additional investment from a player or imitation assumes a permanent effort to monitor others' activity and score their success. These effects can be modeled by considering an additional cost to a specific attitude \cite{rustagi_s10,brede_pone13b,antonioni_pone14,szolnoki_njp14,bertran_jebo17}. As we will show, even a very simple model can provide highly complex behavior and the viability of a certain attitude or strategy updating method depends sensitively on the model parameters. Furthermore, their relation may change repeatedly by varying only a single parameter, but without changing the original character of a certain social dilemma. 

We consider pairwise social games where mutual cooperation provides the reward $R=1$, mutual defection leads to punishment $P=0$. The remaining two payoff values are free parameters of our model to navigate among different dilemma situations. These are the sucker's payoff $S$ of a cooperator against a defector and the temptation value $T$ for the latter player. For simplicity we assume that players are distributed on a square lattice with periodic boundaries where every player interacts with four nearest neighbors when total payoff is calculated. Nevertheless, we stress that our main findings remain unchanged if we use different interaction topologies including triangle and hexagonal lattices or random network.

In addition to the mentioned $C$ and $D$ strategies players are also characterized by a special attitude or trait which determines how they revise their strategies. If a player $x$ is described by the trait imitation ($IM$) then she adopts the strategy $s_y$ from a neighboring $y$ player with a probability 
\begin{equation}
\label{imit}
W (s_x \to s_y) = (1+ \exp[(\Pi_x - \Pi_y)/K])^{-1}\,\,,
\end{equation}
where $\Pi$ denotes the accumulated payoff values gained from two-player games with nearest neighbors. This sum is reduced by an attitude-specific cost of focal player. In particular, an imitating player bears an additional $\epsilon_{IM}$ cost, while a player who uses (myopic) best response ($BR$) to update her strategy should bear $\epsilon_{BR}$. The remaining parameter $K$ determines the noise level of imitation process. In the alternative case, when the $x$ player's attitude is characterized by (myopic) best response to update her strategy, then she changes her $s_x$ strategy to $s'_x$ with a probability 
\begin{equation}
\Gamma (s_x \to s'_x) = (1+ \exp[(\Pi_x - \Pi'_x)/K])^{-1}\,\,,
\end{equation}
where $\Pi_x$ and $\Pi'_x$ are the income of player $x$ when playing $s_x$
and $s'_x$ for the given neighborhood. For simplicity we applied the same noise level as for the above described imitation process. 

Since our principal interest to explore how different attitudes compete we also allow individual attitude to change. When this microscopic process is executed, which is independent from the previously specified strategy update, we assume that a player $y$ forces her attitude or individual trait upon a neighboring player $x$ with the probability defined by Eq.~\ref{imit}. Technically we thus have a four-state model, where strategy and individual attitude coevolve during the evolutionary process. 

\begin{figure}
\begin{center}
\includegraphics[width=8.5cm]{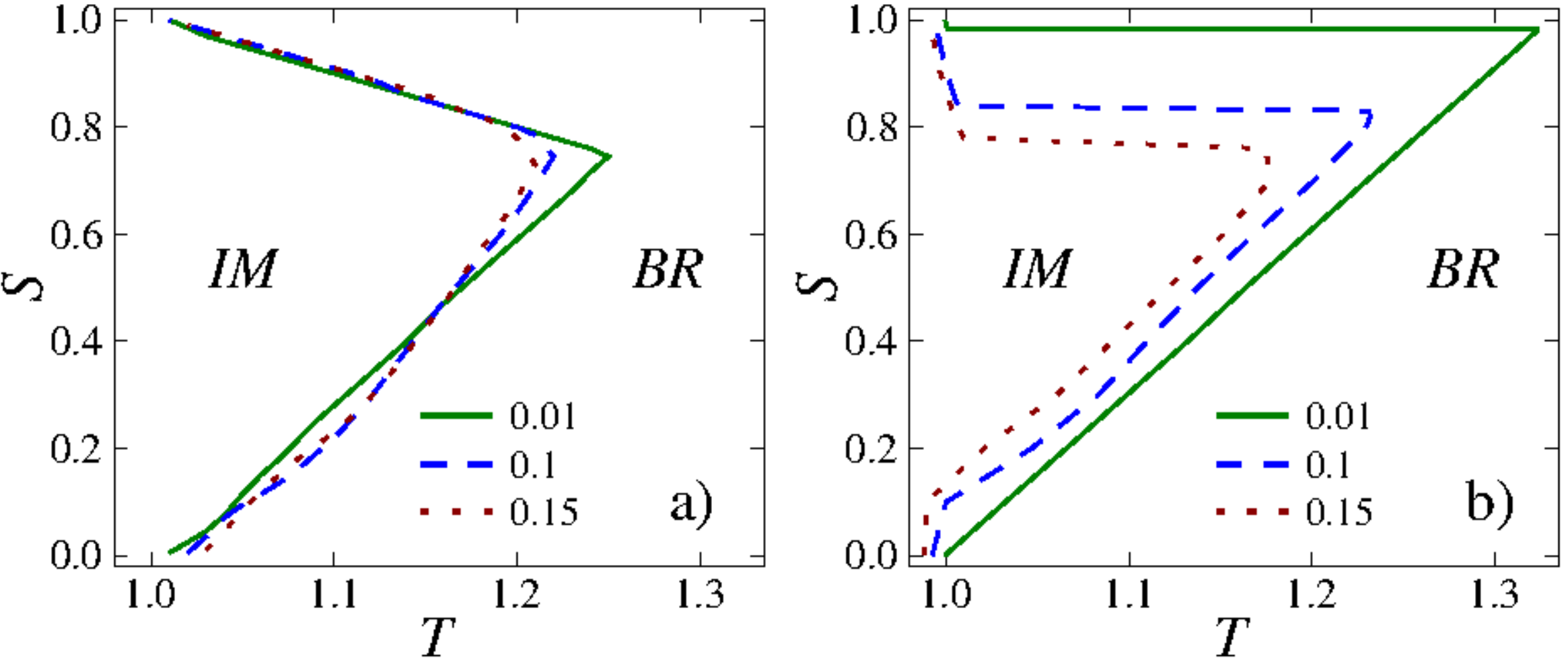}
\caption{\label{no_cost} Border lines on $T-S$ plane of snow-drift game separating the regions where either imitation ($IM$) or best response ($BR$) attitude dominates. Panel~(a) shows the borders calculated from the payoff differences of basic models where either imitation or best response strategy update is used exclusively. Panel~(b) denotes the phase boundaries resulted from the complex model where both attitudes are present in the initial states. $\epsilon_{BR}=\epsilon_{IM}=0$ are used for both panels.
The applied $K$ noise levels are denoted in the legend.}
\end{center}
\end{figure}

We have performed Monte Carlo simulations and monitored the fractions of strategies and attitudes. If players' attitudes reached a uniform state we terminated the simulation because the system become equivalent to a basic model where either imitation or best response rule is used exclusively to update individual strategies \cite{szabo_pre10}. Similarly, if strategy distribution becomes uniform because either $C$ or $D$ strategy goes extinct then we also stopped simulation. In the latter case further evolution becomes uninteresting because in the absence of different strategies the competition of attitudes is determined by their additional costs or, if these are equal, the dynamics resembles to the voter-model like dynamics \cite{cox_ap83,dornic_prl01}. This explains why we only consider snow-drift and stag-hunt games and leave prisoner's dilemma game out. Namely, in the latter case the system practically terminates onto a full defection state and this destination can only be avoided if we assume additional mechanisms \cite{santos_prl05,szolnoki_epl07,perc_pre08}. But the scope of present work is to explore the possible consequence of simultaneous attitudes hence we keep the original basic model without considering further mechanisms.

First we summarize our observations obtained for snow-drift game when no additional costs of attitudes are considered. Figure~\ref{no_cost}(b) highlights that if the $T$ value is close to 1, which means that the temptation to defect is small, then the imitation attitude will spread in the whole system during the coevolutionary process. But for high $T$ temptation values the evolutionary outcome is reversed and the best response attitude crowds out the alternative trait. This observation is in close agreement with the prediction based on the comparison of average payoff values of basic models where only uniform attitude is applied. This comparison is plotted in Fig.~\ref{no_cost}(a) where higher payoff can be reached by applying imitation dynamics at low $T$ values, but best response attitude offers a higher general payoff for individuals when we increase the temptation value.
Interestingly, the payoff difference is practically independent of the applied noise value, but the latter has a significant impact on the phase boundary when attitudes properly compete. As Fig.~\ref{no_cost}(b) shows the higher the noise value the smaller the parameter space where imitation can dominate. This phenomenon can be understood if we consider that the error in imitation will always destroy the efficiency of homogeneous cooperator domains, while this error has no real impact on the role-separating arrangement of $C-D$ pairs when best-response attitude is at work.

\begin{figure}
\begin{center}
\includegraphics[width=8.8cm]{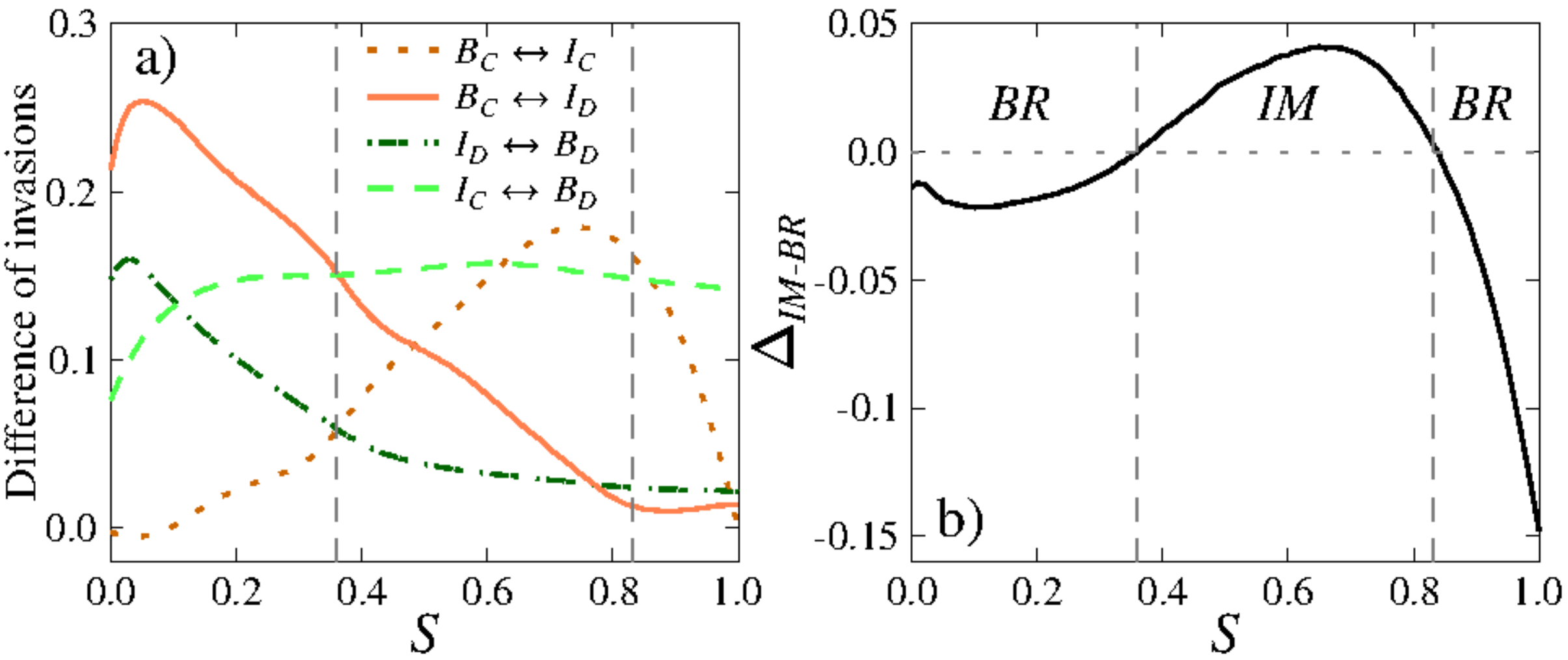}
\caption{\label{T1_1} Efficiency of microscopic invasion processes between different states in dependence on $S$ at fixed $T=1.1$ for $K=0.1$ when no additional costs are considered ($\epsilon_{IM}=\epsilon_{BR}=0$). Only those steps are shown which modify the fractions of competing attitudes. The borders of different phases are marked by dashed vertical lines. While panel~(a) shows the details of specific elementary invasions as described by legend, panel~(b) shows their accumulated values which determine the final outcome of competition. For better clarity we have used $I$ for $IM$ and $B$ for $BR$ players in the legend where elementary invasion processes are specified.}
\end{center}
\end{figure}

Figure~\ref{no_cost}(b) also shows that there is a reentrant phase transition from $BR$ to $IM$ to $BR$ phase as we increase $S$ value at specific fixed $T$ values. This behavior is a straightforward consequence of the relation of cooperator players having different attitudes. At high $S$ the payoff of a cooperator using best response becomes competitive with the payoff of defectors hence the former $B_C$ player can resist the invasion of imitation attitude. Similarly, small positive $S$ value also provides a stable support to $B_C$ players to maintain the checkerboard-like pattern of best response phase. They can resist the invasion of $I_C$ imitator cooperators whose low density in $IM$ phase makes them vulnerable. Between these extreme cases the relatively high $S$ provides a competitive payoff for $I_C$ players whose higher density makes the whole $IM$ phase strong. To confirm this argument in Fig.~\ref{T1_1}(a) we have plotted the differences of elementary invasion steps for all cases where players invade a neighboring site that was occupied previously by a different attitude. This panel shows clearly the non-monotonous change between $I_C$ and $B_C$ states which is mainly responsible for the observed reentrant transition.

\begin{figure}
\begin{center}
\includegraphics[width=8.8cm]{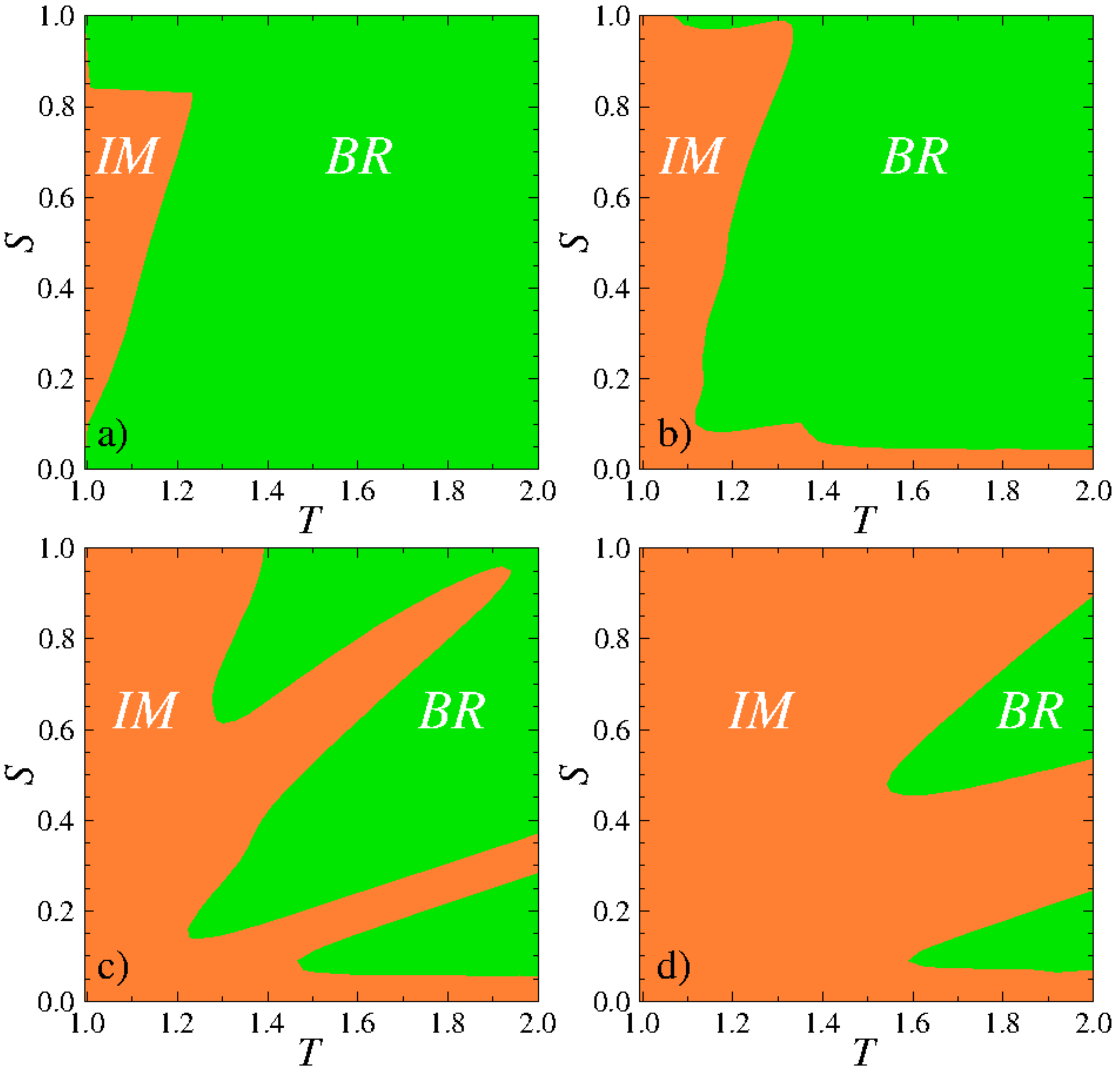}
\caption{\label{phd} Phase diagrams on $T-S$ plane for different cost values of best response update rule while the cost of imitation strategy update remained $\epsilon_{IM}=0$. The former cost is $\epsilon_{BR}=0, 0.05, 0.1,$ and $0.25$ for panel~(a) to panel~(d) respectively. Here orange (green) denotes the parameter area where imitation (myopic best response) attitude prevails as a result of the coevolutionary process. The noise value is $K=0.1$ for all cases.}
\end{center}
\end{figure}

From Fig.~\ref{no_cost}(b) we can conclude that best response attitude can practically dominate the majority of snow-drift quadrant because the emerging role-separating pattern makes it viable. One may expect that if we increase the $\epsilon_{BR}$ cost of this attitude then imitation attitude can gradually invade the whole parameter space. This expectation is justified but in a highly non-trivial way. Figure~\ref{phd} illustrates that the area of $IM$ phase expands as $\epsilon_{BR}$ is increased but the shape of phase separating border could be tangled at intermediate cost values. For example, at $\epsilon_{BR}=0.1, S=0.8$ we can observe three consecutive phase transitions from $IM \to BR \to IM \to BR$ phase by changing only the value of temptation $T$.

\begin{figure}
\begin{center}
\includegraphics[width=8.4cm]{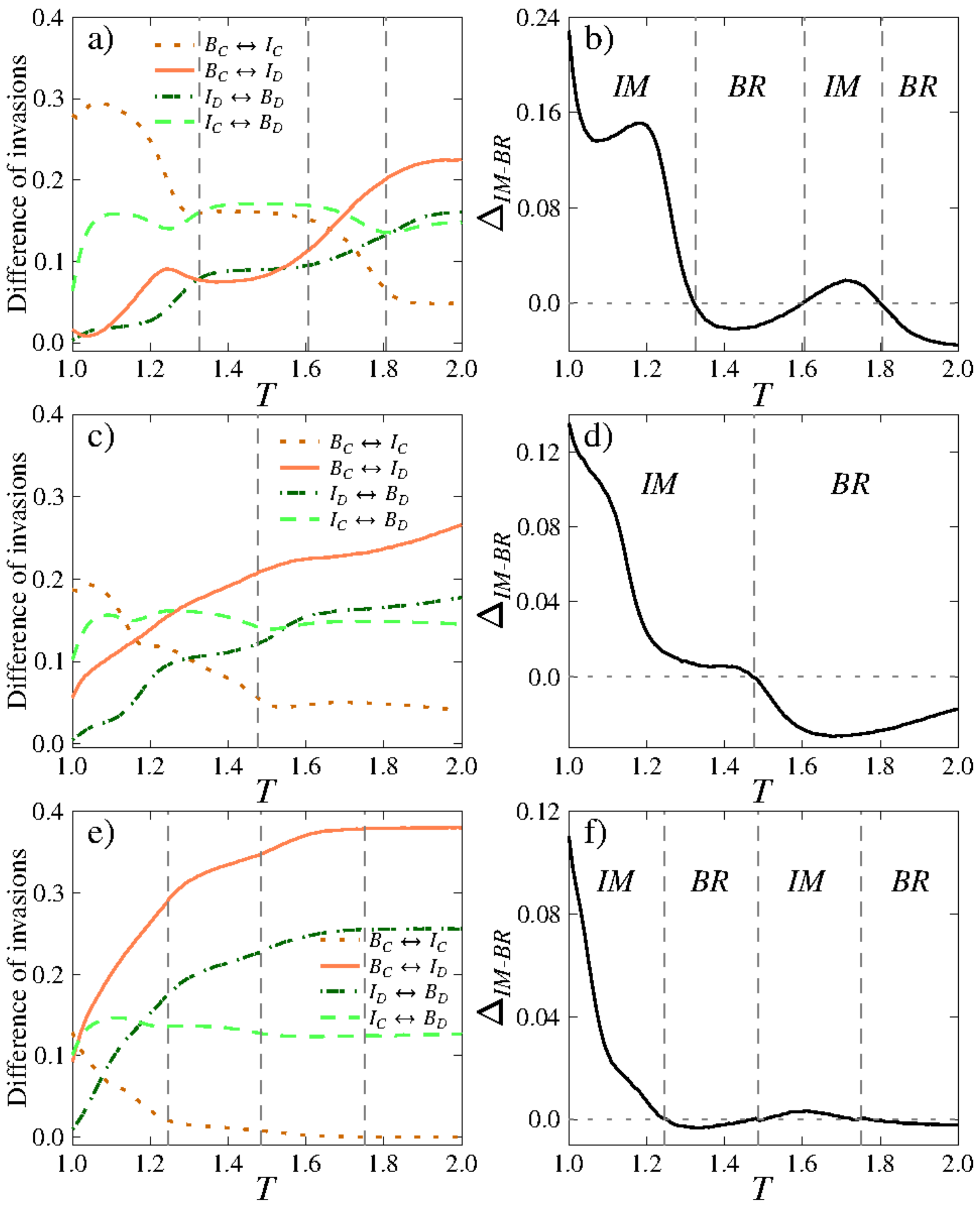}
\caption{\label{rates} The success of elementary invasion steps between competing attitudes for different values of fixed $S$ in dependence on temptation $T$. Top row shows the results for $S=0.8$, middle row for $S=0.5$, and bottom row for $S=0.2$. As for Fig.~\ref{T1_1}, left column shows the full details of invasion, while right column summarizes their impacts on the direction of invasion between competing solutions. As earlier, the critical $T$ values of phase transition points are marked by dashed vertical lines. Other parameters are $K=0.1, \epsilon_{BR}=0.1,$ and $\epsilon_{IM}=0$.}
\end{center}
\end{figure}

In the latter case the explanation of these transitions is more subtle because it cannot be confirmed by comparing only a single pair of competing states. As earlier, in Fig.~\ref{rates} we have recorded the successful elementary invasion steps at three representative $S$ values in dependence on $T$. The explanation of three transitions at high $S$ value, shown in top row, is the following. If we start increasing temptation from $T=1$ then $I_D$ becomes more powerful and simultaneously $I_C$ weakens. At the same time $B_C$ remains intact in the $BR$ domain because $S$ remains high. As a result, $I_C$ weakens against $B_C$ which involves the decay of $IM$ phase against $BR$ phase. Indeed, $I_D$ becomes also stronger against $B_C$, but the former effect is more substantial, as Fig.~\ref{rates}(b) panel illustrates. Increasing $T$ further the average cooperation level does not change relevantly. (This plateau was illustrated in Fig.~4(c) of Ref.\cite{szabo_pre10} where the basic $IM$ model was studied.) However, the further increase of $T$ makes $I_D$ even powerful. As a result, $I_D$ can invade $B_C$ more intensively, which will reverse the direction of propagation between $BR$ and $IM$ phases. The last transition can be explained by the relation of $B_D$ and $I_D$ players, which becomes important for this parameter region. While the former remains fit among $B_C$ players the latter cannot utilize high $T$ because the density of $I_C$ players decays rapidly. This is why $B_D$ will beat $I_D$ more frequently which causes the victory of $BR$ phase again. 

At intermediate $S$ value, shown in middle row of Fig.~\ref{rates}, the previously mentioned plateau of the basic $IM$ model disappears, hence $I_D$ players are unable to utilize the constant support of $I_C$ neighbors. Consequently, we can observe only a single transition from $IM$ to $BR$ phase. At smaller $S$ value, however, we face a new situation because small $S$ cannot maintain $I_C$ players in $IM$ phase for higher $T$ values. This is illustrated by the invasion rates shown in the bottom row of Fig.~\ref{rates} where the invasion success of $B_C$ against $I_C$ diminishes for $T>1.5$. Here $B_C$ cannot beat $I_C$ players anymore and the advantage of $BR$ phase over $IM$ solution disappears. Instead, a pure $I_D$ phase competes with the previously mentioned checkerboard-like pattern of $BR$ phase. Here $I_D$ players can utilize their advantage over $B_D$ players who have to bear the extra $\epsilon_{BR}=0.1$ cost. As a result, $IM$ phase strikes back when temptation exceeds $T=1.5$ value. As we increase $T$ further, the disadvantage of additional cost becomes marginal and the stable support of $B_C$ neighbors will provide a competitive payoff for $B_D$ players, which explains why $BR$ phase can win again.

\begin{figure*}
\begin{center}
\includegraphics[width=14.5cm]{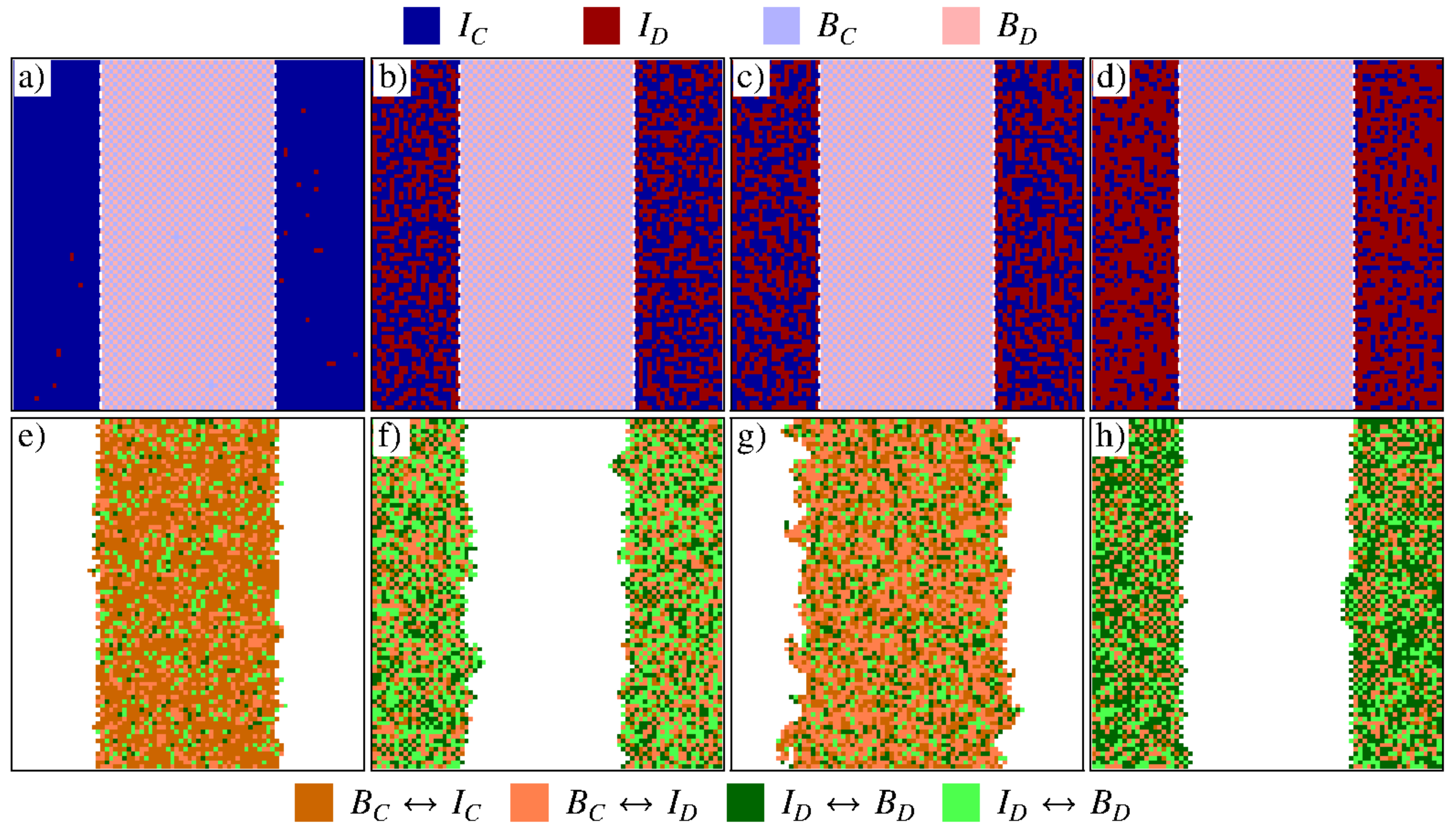}
\caption{\label{trace} Competition of attitudes for different temptation values for $T=1.15, 1.45, 1.7,$ and $1.9$ (from left to right). The other parameters, $S=0.8, \epsilon_{IM}=0, \epsilon_{BR}=0.1, K=0.1$ and $L=80$, are fixed for all cases. Top row shows the initial separation of lattice where the composition of best response players are surrounded by players who apply imitation-based strategy update rule. First invasion across the vertical phase separating lines are forbidden hence subsystem solutions are relaxed to the characteristic states which are determined by $T, S,$ and $K$ values. When competition starts by removing the border between them then either $IM$ or $BR$ phase prevails depending on the $T$ value (not shown). Bottom row illustrates the trace of elementary processes during the invasion. To distinguish them we used the same color coding as for Fig.~\ref{T1_1} and Fig.~\ref{rates}.}
\end{center}
\end{figure*}

The comparative plots of Fig.~\ref{trace} provide a deeper insight into the consecutive phase transitions as we increase the temptation value. Here we first separated the lattice into two parts where the solutions of basic models evolved independently due to the applied parameter values. More precisely, players using best response attitude were closed in the central domain where this subsystem relaxed to the $BR$ phase, while players using imitation attitude were in the surrounding space where $IM$ phase evolved.
In other words, neither strategy nor attitude transfer was allowed across the separating borders which are marked by dashed white lines. These final states of the relaxation, which are the initial states of attitude competitions, are plotted in the top row of Fig.\ref{trace}. After we removed the borders, the starting strategy and attitude transfer resulted in a complete success of one of the basic solutions. We note that the final states are not shown here, but can be read out from the top row of Fig.\ref{rates}. Instead, we have recorded the "trace" of invasion steps for every cases. More precisely in the bottom row of Fig.~\ref{trace} we colored those lattice sites where invasion happened during the whole competition until sole $IM$ or $BR$ state was reached. The applied colors, which are plotted in the bottom of the Figure, mark the last invasion process at a given position. 

Figure~\ref{trace}(a) demonstrates that at small $T$ value the $IM$ state is full of $I_C$ players who can support each other effectively and collect high payoff value. As a consequence, $IM$ phase can easily invade $BR$ phase at this parameter region. The corresponding Fig.~\ref{trace}(e) illustrates that in this case the most typical change between the competing states is when the previously mention strong $I_C$ player invades the weaker member of $BR$ phase, which is $B_C$ player. As we increase temptation value, shown in Fig.~\ref{trace}(b), the density of $I_C$ players decays which weakens them significantly. At the same time $I_D$ cannot gain enough power because the $T$ value is still moderate. As a result, the direction of invasion turns back and $BR$ starts propagating. Indeed, the related Fig.~\ref{trace}(f) demonstrates that the $B_D \to I_D$ and $B_D \to I_C$ transitions become dominant. As we already noted, by increasing $T$ further the density of $I_C$ players does not change relevantly due to the high value of $S$. This is clearly visible in Fig.~\ref{trace}(c), where $IM$ phase before the competition remained practically unchanged. It means that $I_D$ players can enjoy undisturbed support from $I_C$ neighbors but the former is already armed by a higher $T$ payoff. That explains why the $IM$ phase can invade again because the $I_D \to B_C$ transition, marked by light orange, becomes relevant. Lastly, if we increase temptation value $T$ further then $I_D$ becomes too successful within $IM$ phase, hence the density of $I_C$ players decays drastically, as it is shown in Fig.~\ref{trace}(d). Consequently, $I_D$ players are unable to enjoy the support of neighboring $I_C$ players when they fight against the external $BR$ phase. $B_D$ players of the latter phase, however, can still enjoy the solid support of $B_C$ neighbors due to the checkerboard-like patter of this phase. That explains why $B_D$ players can beat $I_D$ players, and $BD$ phase invades $IM$ phase no mater the former attitude should still bear an extra cost. This phenomenon is nicely illustrated in Fig.~\ref{trace}(h) where dark green pixels emerged more frequently. To summarize the surprisingly different outcomes of evolution processes we have provided an animation \cite{S08}, where all discussed cases are shown simultaneously using the same $S=0.8, \epsilon_{BR}=0.1$ values and the only difference is the temptation value as it is described by Fig.~\ref{trace}.

\begin{figure}
\begin{center}
\includegraphics[width=8.5cm]{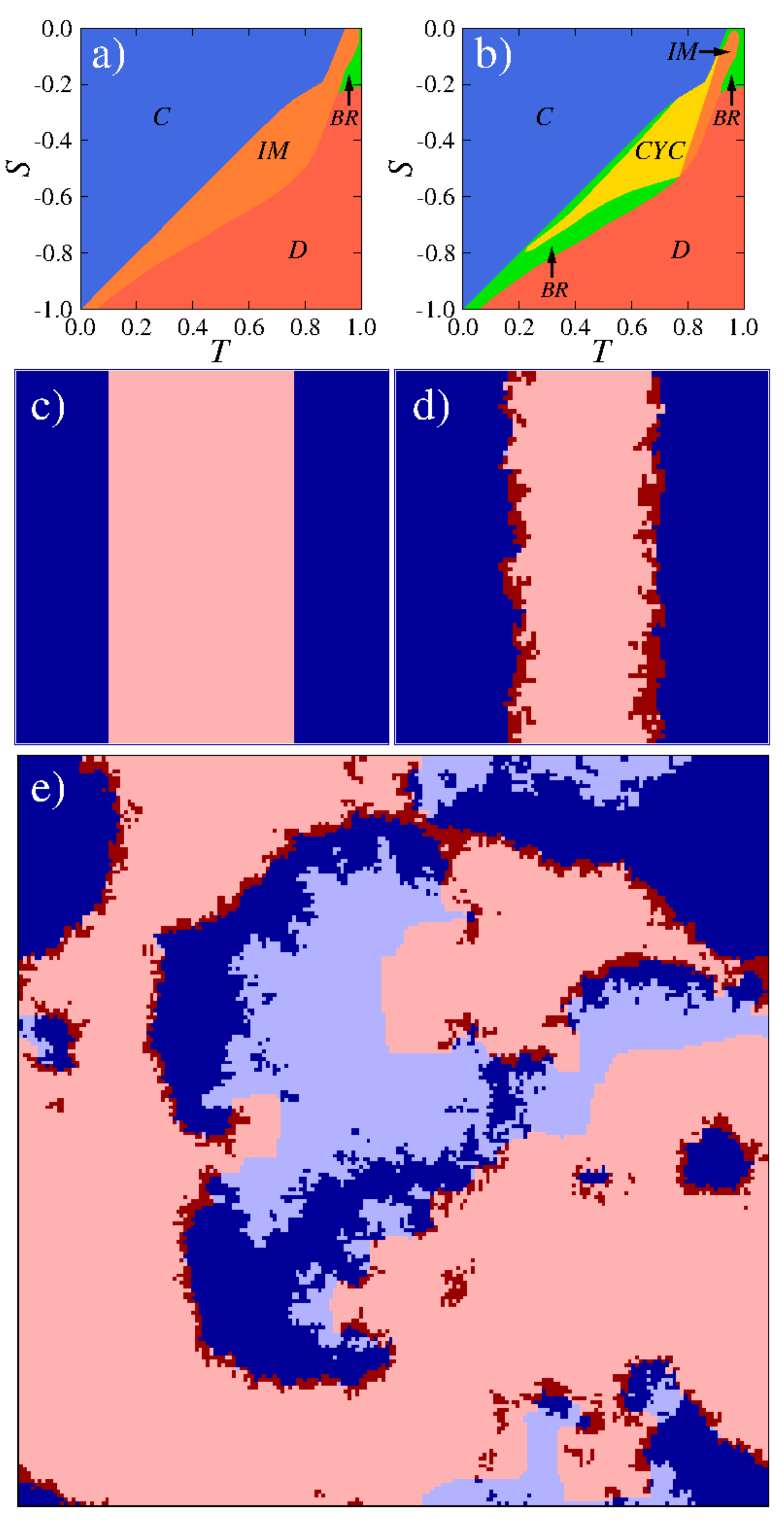}
\caption{\label{SH} Panel~(a): phase diagram on $T-S$ plane of stag-hunt game for $\epsilon_{BR}=\epsilon_{IM}=0$ at $K=0.1$.
$C$ ($D$) denotes the phase where defectors (cooperators) die out very early. When $IM$ imitation attitude prevails, the system evolves into a full cooperator state.
Panel~(b) shows the phase diagram when $\epsilon_{IM}=0.02$ is applied for imitation attitude. Here $CYC$ denotes a solution where all available states coexist due to cyclic dominance.
Panel~(c) shows the initial state of competition when stable solutions of basic models (full $I_C$ and full $B_D$) start competing at $T=0.35, S=-0.7$, and $\epsilon_{IM}=\epsilon_{BR}=0$. Panel~(d) illustrates a representative intermediate state of invasion before imitation attitude invades the whole system. During the invasion a new state, $I_D$, emerges which highlights the unequal invasion speeds of attitude and strategy. Panel~(e) illustrates the cyclic dominance between $I_C \to B_D \to B_C \to I_C$ states at $T=0.7, S=-0.5$, and $\epsilon_{IM}=0.02$.
The color codes of microscopic states are identical to those we used in top row of Fig.~\ref{trace}.}
\end{center}
\end{figure}

In the rest of this work we present our observations obtained for stag hunt game, where $R > T > P> S$ rank characterizes the dilemma. The most fundamental difference from the above discussed snow-drift dilemma is best response attitude cannot provide a checkerboard-like pattern here, hence homogeneous solutions compete for space \cite{szabo_jtb12}. In this situation imitation is more effective when both attitudes are free from additional cost, because $IM$ attitude can extend full $C$ state to a larger area on $T-S$ plane. This is illustrated in Fig.~\ref{SH}(a), where we plotted the phase diagram using $\epsilon_{IM}=\epsilon_{BR}=0$ cost values at $K=0.1$. If $T$ is too small then defectors die out very early and both basic models terminate into a full cooperator state. Increasing $T$ best response attitude does better and invades the whole space. This state is marked by $IM$, but we note that full cooperator state is still maintained. Increasing temptation further there is a sharp transition into the full $D$ state that is in agreement with the basic models where uniform attitudes are assumed \cite{szabo_pre10}. 

The invasion of $IM$ phase into the $BR$ phase reveals an interesting phenomenon that is based on the unequal propagation speeds of strategy and attitude. To illustrate it in Fig.~\ref{SH}(c) we start the evolution from an initial state where two stable solutions of basic models are present at $T=0.35, S=-0.7$. More precisely $B_D$ players, who are in the middle of this panel are fighting against $I_C$ players who surround them. When evolution starts $B_D$ players at the frontier change their attitude first and become $I_D$ players. This new state, which is not present in the initial state, is marked by dark red color in Fig.~\ref{SH}(d). The whole propagation process can be followed in an animation we provided as supplementary information \cite{Moor}. It is important to note that this new state has a special role on the propagation of $I_C$ players. On one hand, $I_D$ cannot be utilized by $M_D$ players, but on the other hand the former could be more successful than the latter since they enjoy the vicinity of $I_C$ players. This explains why $I_D$ (dark red) propagates in the sea of $B_D$ (light red). Interestingly, the triumph of $I_D$ is just temporary because they immediately are invaded by $I_C$ players. The latter process ensures a thin protecting skin around $I_C$ domain in a self-regulating way. Put differently, $I_D$ helps $I_C$ to invade $BR$ phase and after, fulfilling its job, $I_D$ goes extinct. This is the so-called "the Moor has done his duty, the Moor may go" effect which was previously observed in a completely different system where punishing strategies were involved in a public goods game \cite{szolnoki_pre11b}.

Naturally, if we apply a significant cost for imitation attitude then it looses its advantage and players using best response attitude will dominate. As a result, the area of full $C$ state shrinks on the $T-S$ plane and its border shifts to the $S=T-1$ line in the zero noise limit, which characterizes the $BR$ basic model. Interestingly, a moderate $\epsilon_{IM}$ cost allows a new kind of solution to emerge. To illustrate it we present a phase diagram plotted in Fig.~\ref{SH}(b) where $\epsilon_{IM}=0.02$ was applied. This diagram suggests that at some parameter values all competing states can survive and coexist. This coexistence is based on a cyclic dominance between microscopic states and a typical spatial pattern is plotted in Fig.~\ref{SH}(e). As the pattern suggests $I_C$ (dark blue) invades $B_D$ (light red) with the help of $I_D$ (dark red) players. Here the role of $I_D$ is the same as we described above. However, $B_C$ (light blue) invades $I_C$ (dark blue) because the former should bear an extra cost. Lastly, $B_D$ (light red) invades $B_C$ (light blue) because the best response basic model dictates a full $D$ state at this $T-S$ parameter values. For clarity we also provided an animation where the dynamics of this states can be followed \cite{cyc}.

The above description of cyclic dominance explains why we cannot observe coexistence for too high $\epsilon_{IM}$ values. In the latter case the vicinity of $I_C$ cannot compensate the high cost value of $I_D$, hence $I_D$ cannot invade $B_D$ domain anymore. As a result, the cyclic chain of invasions is broken and system terminates into a state where the population is described by a homogeneous state. This behavior is in close agreement with our general understanding about the positive role of cyclic dominance to maintain diversity of microscopic states \cite{kerr_n02,szolnoki_jrsif14,rulquin_pre14,reichenbach_n07,avelino_pre14,roman_jtb16,perez_srep16,lutz_g17,bazeia_epl17,dobramysl_jpa18}.

To sum up, we have shown that the success of different strategy updating traits or attitudes may depend sensitively on the actual payoff values which characterize a social dilemma. In most of the parameter regions we detected homogeneous populations but we can observe several transitions between an imitation dominant state to a population which is described by best response attitude. Our key finding is attitudes and strategies may propagate with different speeds which makes it possible for several interesting pattern formation to emerge. For example, consecutive re-entrant phase transitions are detected by only changing a single parameter without modifying the fundamental character of a social dilemma. We have also shown that cyclic dominance can emerge between microscopic states no matter there are only two major $C$ and $D$ strategies. Indeed, it was previously found that a two-strategy system can produce similar cyclic dominance in spatial systems \cite{szolnoki_pre10b}, but the mentioned example assumed diverse timescales during the evolution. Our present observations emphasize that the microscopic origin of diversity has just a second-order importance because every type of microscopical diversity could be a source of cyclical dominance among competing states. 

We note that all the presented results are robust to replacing lattice-type interaction topology by random graph, and can be observed also for other parameter values. We conclude that considering simultaneous presence of different strategy updating or learning attitudes might be a new research avenue for modeling human behavior in social dilemmas more realistically.

\begin{acknowledgments}
This research was supported by the Hungarian National Research Fund (Grant K-101490) and Natural Science Foundation of Zhejiang Province (Grant Nos. LY18F030007 and LY18F020017).
\end{acknowledgments}

\end{document}